\documentstyle[12pt,epsf]{article}
\makeatletter
\renewcommand{\theequation}{\thesection.\arabic{equation}}
\@addtoreset{equation}{section}
\makeatother

\begin{document}

\rightline {LAPTH-701/98}

\rightline{hep-ph/9810273}

\vspace*{0.5cm}

\begin{center}

{\bf THE HIGGS SECTOR OF THE MINIMAL 3 3 1 MODEL  REVISITED}
\vspace*{1cm}

{\bf Nguyen Tuan Anh $^a$}, 
~ {\bf Nguyen Anh Ky $^{b,c}$}  and  {\bf Hoang Ngoc Long $^{b,d }$
\footnote{Fellow of the Japan Society for the Promotion of Science}}
\vspace*{0.5cm}

$^a$ 
{\it Graduate School, Institute of Physics, P. O. Box
429, Bo Ho, Hanoi 10000, Vietnam}\\

$^b$ {\it Institute of Physics, NCNST, P. O. Box
429, Bo Ho, Hanoi 10000, Vietnam}\\

$^c$ {\it LAPTH, Chemin de Bellevue, B.P. 110, F--74941 Annecy--le--Vieux, Cedex,
France}\\

$^d${\it Department of Physics, Chuo University, Kasuga, Bunkyo-ku,
Tokyo 112-8551, Japan}\\
\vspace*{1cm}

{\bf Abstract}
\end{center}
The mass spectrum and  the eigenstates of the Higgs sector
of the minimal 3 3 1 model are revisited in detail.
There are discrepancies
%in
between our results and  previous results by another author.

\vskip 1 true cm

\newpage

\section{Introduction}
\label{intro}

\hspace*{0.5cm}At present, most high energy physicists 
recognize the great success
of the Standard Model (SM) of electroweak and strong interactions.
%$\mbox{SU}(2)_L \otimes \mbox{U}(1)$ model.
However, according to our present knowledge, the model is able
to describe only the phenomenology at low energy below 
200 GeV. In addition, one of the basic elements of the
model is still an open problem, namely, the observation of the Higgs boson.
Hence the mechanism for electroweak symmetry breaking is, in some way, 
still a mystery. The scalar sector  plays an important  role in 
many subjects of physics,
and one of the most urgent problems in high energy physics
is the search for the Higgs bosons.
The scalar sector has been thoroughly studied not only in the
SM framework but also in its various extensions.
The models based on the $\mbox{SU}(3)_C\otimes \mbox{SU}(3)_L \otimes 
\mbox{U}(1)_N$ (3 3 1) gauge
group [1--6] is one of the most interesting extensions.
These models have the following intriguing features:
firstly, the models are anomaly free only if the number
of families N is a multiple of three. Further,
from the condition of QCD asymptotic freedom,
which is valid only if the number of quark families
is to be less than five, it follows that N
is equal to 3. The second characteristic
is that  the Peccei-Quinn (PQ)~\cite{pq} symmetry -- a solution
of the strong CP problem  naturally occurs in these models~\cite{pal}.
It is worth mentioning that
the implementation of the PQ symmetry
is usually possible only at classical level (it will be  broken
by the quantum corrections through instanton effects),
and there has been a number of attempts to find models
for solving the strong CP question. In these models the PQ symmetry
following from the gauge invariant Lagrangian  does not
have to be imposed. The third interesting feature
is that one of the quark families  is treated differently
from the other two. This could lead to a natural explanation
for the unbalancing heavy top quarks, deviations of $A_b$
from the SM prediction etc.  In addition, the models
predict no very high new mass scales at the order of 
a few TeV~\cite{dng}.\par
\bigskip
The scalar sector of the minimal 3 3 1 model was studied recently by
Tonasse~\cite{ton}. Three Higgs triplets were  firstly  analysed,
then the sextet was added  in a further consideration.
Unfortunately, the last case -- the three triplet and one sextet
was presented so briefly such that the reader could hardly understand
how it was  obtained. Comparing with our results, besides some
minor mistakes  (misprints, we guess) in~\cite{ton}, we also
found several discrepancies in the multiplet structures
and some of the conclusions, especially those from the graphic analysis.
In the present paper, considering again the Higgs sector we correct and
present the results in a systematic order so that they  can be checked
and used in  further studies.\par
\bigskip
The Higgs potentials of the model, constraint equations and
main notations are  presented in  section  {\bf 2}.
Sec. {\bf 3} is devoted to solving  the characteristic equations.
Our results are summarised in the last section. The full expression of 
the potential with three triplets and one sextet is given in the Appendix A,
while the results for the toy model -- the three triplet model is presented 
in the Appendix B where we explain why our results differ from those
obtained in \cite{ton}.

\section{Higgs structure and potential}

In the original version of the 3 3 1 model~\cite{ppf} the Higgs sector
consists of three triplets
\begin{equation}
\eta = \begin{array}{cccccc}
\left(\begin{array}{c}
\eta^0 \\ \eta^-_1 \\ \eta^+_2
\end{array}\right) \sim (1,3,0); \hspace*{0.2cm}
\rho = \left(\begin{array}{c}
\rho^+ \\ \rho^0 \\ \rho^{++}
\end{array}\right) \sim (1,3,1); \hspace*{0.2cm}
\chi = \left(\begin{array}{c}
\chi^- \\ \chi^{--} \\ \chi^0\end{array}\right) \sim (1,3,-1),
\end{array}
\label{hig1}
\end{equation}
where the numbers in the brackets denote the quantum numbers under
$\mbox{SU}(3)_C, \mbox{SU(3)}_L$ and $\mbox{U(1)}_N$ respectively.\par
The most general (i.e., renormalizable and gauge invariant)
SU(3)$_L\otimes\-$U(1)$_N$ Higgs potential which we can write
with the three triplets of the Eqs. (\ref{hig1}) is given by
\begin{eqnarray}
V_T(\eta,\rho,\chi)& = &\mu_1^2\eta^\dagger\eta+\mu^2_2\rho^\dagger\rho +
\mu^2_3\chi^\dagger\chi+
\lambda_1(\eta^\dagger\eta)^2 + \lambda_2(\rho^\dagger\rho)^2 +
\lambda_3(\chi^\dagger\chi)^2
\nonumber
\\ &+&\mbox{}
(\eta^\dagger\eta)\left[\lambda_4(\rho^\dagger\rho )+\lambda_5(\chi^\dagger
\chi)\right] +\lambda_6(\rho^\dagger\rho)(\chi^\dagger\chi) +
\lambda_7(\rho^\dagger\eta)(\eta^\dagger\rho) \nonumber \\
&+ &\mbox{}  \lambda_8(\chi^\dagger\eta)(\eta^\dagger\chi) +
\lambda_9(\rho^\dagger\chi)(\chi^\dagger\rho) +
\left(\frac{f_1}{2}\varepsilon^{ijk}\eta_i\rho_j\chi_k + \mbox{h. c.}\right),
\label{potl1}
\end{eqnarray}
where the $\mu$'s,  and $f_1$ are mass parameters and
coupling constants having a dimension of mass, while $\lambda$'s are
dimensionless.
% Comparing our results with those in 
%Ref.~\cite{ton}, besides several minior mistakes
%(misprints, we guess) in ~\cite{ton} there are 
%discrepancies in some of the multiplet structures 
%and conclusions. 
The detailed analysis of this
potential is given in Appendix B where we explain 
the reason why our results obtained here differ 
from those by Tonasse in \cite{ton}.\par  

  It was found soon after that  in this model not all of the leptons got
a mass, and this problem was solved by adding to the scalar sector
a sextet~\cite{rf}
\begin{equation}
S=\left(\begin{array}{ccc}
\sigma^0_1 & s_2^+ & s_1^- \\
s_2^+ & s_1^{++} & \sigma_2^0 \\
s_1^- &\sigma_2^0 & s_2^{--}
\end{array}\right) \sim (1,\bar{6}, 0).
\label{hig2}
\end{equation}
To avoid  unwanted terms which make the analysis of the Higgs sector
more complicated and lead to nonzero Majorana neutrino masses,
a discrete symmetry should be imposed (for details, see~\cite{rf}).
Thus, we have additional terms in the Higgs potential in
Eq. (\ref{potl1}). The new (modified) potential is
\begin{eqnarray}
V_S\left(\eta , \rho , \chi, S\right) & = & V_T + \mu^2_4
\mbox{Tr}\left(S^\dagger
S\right) + \lambda_{10}\mbox{Tr}^2\left(S^\dagger S\right) +
\lambda_{11}\mbox{Tr}\left[\left(S^\dagger S\right)^2\right]  \cr
&+&  \left[\lambda_{12}\left(\eta^\dagger\eta\right) +
\lambda_{13}\left(\rho^\dagger\rho\right) +
\lambda_{14}\left(\chi^\dagger\chi\right)\right]
\mbox{Tr}\left(S^\dagger S\right)  \cr
&+&  \frac{1}{2}\left(f_2\rho^T S \chi + \mbox{h.c.}\right),
\label{pot}
\end{eqnarray}
where $V_T$ is given in Eq. (\ref{potl1}).
Let the neutral scalars $ \eta^0, \rho^0, \chi^0$, and $\sigma_2^0$
develop {\it real} vacuum expectation values (VEV's) $v$, $u$, $w$ and $v'$,
respectively. (The CP--phenomenology arising from complex vevs in the 331 
models has been inverstigated in \cite{gum}). 
We rewrite the expansion of the scalar fields
\begin{equation}
\eta^0 = v + \xi_\eta + i\zeta_\eta ; \
\rho^0 = u + \xi_\rho + i \zeta_\rho;\
\chi^0 = w + \xi_\chi + i \zeta_\chi;\
\sigma^0_2 = v' + \xi_\sigma + i \zeta_\sigma.
\label{exp1}
\end{equation}
As in~\cite{rf} here we do not consider neutrino mass, hence
$\sigma_1^0$ does not develop a vacuum expectation
\begin{equation}
\sigma_1^0 = \xi'_\sigma + i \zeta'_\sigma.
\label{exp2}
\end{equation}
Following \cite{ic97} we call a real part $\xi$ scalar and an imaginary one
$ \zeta$ -- pseudoscalar.
The pattern of the symmetry breaking is\par
\medskip
\centerline{$\mbox{SU(3)}_C\ \otimes \
\mbox{SU(3)}_L\otimes \mbox{U(1)}_N
\stackrel{\langle\chi\rangle}{\longmapsto}
\mbox{SU(3)}_C \ \otimes \ \mbox{SU(2)}_L\otimes
\mbox{U(1)}_Y \stackrel{\langle\rho,\eta, S\rangle}{\longmapsto}
\mbox{SU(3)}_C \ \otimes \ \mbox{U(1)}_Q$,}
\medskip\par
\noindent and the VEV's are related to the standard model one ($v_W$) as
$v^2 + u^2 + v'^2 = v_W^2$ = (246 Gev)$^2$.\par
At the first step of symmetry breaking, the large  $\langle \chi \rangle$
will generate masses for exotic quarks and new heavy gauge bosons  $Z',
X^{\pm \pm}, Y^{\pm}$.
The subsequent breaking of $ \mbox{SU(2)}_L \otimes \mbox{U(1)}$ is
accomplished with nonzero values of $\langle \rho \rangle$ and
$\langle \eta \rangle$ ,
such that $t, s$ and $d$ aquire masses proportional to the former, while
$b,  c$, and $u$ aquire masses proportional to the latter.\par
To keep  the model consistent with low--energy phenomenology,
the VEV  $\langle \chi \rangle$ must be large enough.
In this paper we will use the following assumption: VEV of the Higgs
field at the first step of symmetry breaking is assumed to be much
larger than those at the second step, i.e.,
\begin{equation}
w \gg v, u, v'.
\label{lecod}
\end{equation}
For further use we write
\[Tr(S^\dagger S)=2(v'^2+2v'\xi_\sigma+\xi^2_\sigma+\zeta^2_\sigma)+
\xi'^2_\sigma+\zeta'^2_\sigma+2s_1^+s_1^-+2 s_2^+ s_2^-+
s_1^{++}s_1^{--}+s_2^{++}s_2^{--},\]
\begin{eqnarray}
Tr[(S^\dagger S)^2]&=&(\xi'^2_\sigma+\zeta'^2_\sigma+
s_1^+s_1^-+s_2^+s_2^-)^2+
(v'^2+2 v'\xi_\sigma+\xi^2_\sigma+\zeta^2_\sigma+
s_2^+s_2^-+s_1^{++}s_1^{--})^2\nonumber\\
&+&(v'^2+2v'\xi_\sigma+\xi^2_\sigma+\zeta^2_\sigma+
s_1^+s_1^-+s_2^{++}s_2^{--})^2\nonumber\\
&+&2[(v'^2+2 v'\xi_\sigma+\xi^2_\sigma+\zeta^2_\sigma)
s_1^+s_1^-+(\xi'^2_\sigma+\zeta'^2_\sigma)s_2^+s_2^-+
\sigma_1^o\sigma_2^os_1^+s_2^-+
\sigma_1^{o*}\sigma_2^{o*}s_2^+s_1^-\nonumber\\
&+&\sigma_1^{o*}s_2^+s_2^+s_1^{--}+
\sigma_1^{o}s_2^-s_2^-s_1^{++}+
\sigma_2^{o*}s_1^-s_2^-s_1^{++}+
\sigma_2^os_1^+s_2^+s_1^{--}+
s_2^+s_2^-s_1^{++}s_1^{--}]\nonumber\\
&+&2[(v'^2+2v'\xi_\sigma+\xi^2_\sigma+\zeta^2_\sigma)
s_2^+s_2^-+(\xi'^2_\sigma+\zeta'^2_\sigma)s_1^+s_1^-+
\sigma_1^o\sigma_2^os_1^+s_2^-+
\sigma_1^{o*}\sigma_2^{o*}s_2^+s_1^-\nonumber\\
&+&\sigma_1^{o^*}s_1^-s_1^-s_2^{++}+
\sigma_1^{o}s_1^+s_1^+s_2^{--}+
\sigma_2^{o*}s_1^+s_2^+s_2^{--}+
\sigma_2^os_1^-s_2^-s_2^{++}+
s_1^+s_1^-s_2^{++}s_2^{--}] \nonumber\\
&+&2[(v'^2+2 v'\xi_\sigma+\xi^2_\sigma+\zeta^2_\sigma)
(s_1^{++}s_1^{--}+s_2^{++}s_2^{--})+
\sigma_2^o\sigma_2^os_2^{++}s_1^{--}+
\sigma_2^{o*}\sigma_2^{o*}s_1^{++}s_2^{--} \nonumber\\
&+&\sigma_2^os_1^-s_2^-s_2^{++}+
\sigma^o_2s^+_1s^+_2s^{--}_1+
\sigma_2^{o*}s_1^+s_2^+s_2^{--}
+\sigma_2^{o*}s_1^-s_2^-s_1^{++}
+s_1^+s_1^-s_2^{+}s_2^{-}],
\label{bps}
\end{eqnarray}
and the complete expression of $V_S$ is given in the Appendix A.\par

The requirement that in the shifted potential $V_S$, the linear terms in fields
must be absent, gives us the following constraint equations
in the tree level approximation
\begin{eqnarray}
\mu^2_1 + 2 \lambda_1 v^2 + \lambda_4 u^2 + \lambda_5 w^2 +
2 \lambda_{12} v'^2 + \frac{f_1 u w}{2 v}
 & = & 0,\nonumber\\
\mu^2_2 + 2 \lambda_2 u^2 + \lambda_4 v^2 + \lambda_6 w^2 +
2 \lambda_{13} v'^2 + \frac{f_1 v w}{2 u} +  \frac{f_2 v' w}{2 u}
 & = & 0,\label{cont}\\
\mu^2_3 + 2 \lambda_3  w^2 + \lambda_5 v^2 + \lambda_6 u^2 +
2 \lambda_{14} v'^2 + \frac{f_1 v u}{2 w} +  \frac{f_2 u v'}{2 w}
 & = & 0,\nonumber\\
\mu^2_4 + 2 (2 \lambda_{10}  + \lambda_{11}) v'^2 +
\lambda_{12} v^2 + \lambda_{13} u^2 +
\lambda_{14} w^2 + \frac{f_2 u w}{4 v'}
 & = & 0\nonumber.
\end{eqnarray}
Substituting  Eqs.~(\ref{hig1}),~(\ref{hig2}),~(\ref{exp1})
and~(\ref{exp2}) into~(\ref{potl1}) and~(\ref{pot}), and diagonalizing,
 we will get a mass spectrum of
Higgs bosons with mixings.

\section{Higgs eigenstates and mass spectrum}

Since $\sigma_1^0$ has not a VEV, the associated scalar $\xi'_\sigma$ and
pseudo--scalar $\zeta'_\sigma$ do not mix with other fields (for details,
see~\cite{ic97}) and we indeed have the physical field \footnote {Here we
keep the notations in~\cite{ton} except the subscript 0 in eigenstates is
omitted.}
  $H'_\sigma \approx
\xi'_\sigma$ with mass
\begin{equation}
m^2_{H'_\sigma}= 2\lambda_{11}v'^2 +
\frac{f_2 u w}{4v'}.
\end{equation}
In the $\xi_\eta, \xi_\rho, \xi_\sigma$  and $ \xi_\chi$ basis
the square mass matrix, after imposing the constraints
~(\ref{cont}), reads as
\begin{equation}
M^2_{4\xi} = \left(
\begin{array}{cccc}
4\lambda_1v^2-\frac{f_1uw}{2v}&2\lambda_4vu+\frac{f_1w}{2}
&4\lambda_{12}vv'& 2\lambda_5vw+\frac{f_1u}{2} \\
2\lambda_4vu+\frac{f_1w}{2}&4\lambda_2u^2-\frac{w}{2u}(f_1v+f_2v')
&4\lambda_{13}uv'+\frac{f_2w}{2}&
 2\lambda_6uw + \frac{f_1v}{2} +\frac{f_2v'}{2} \\
 4\lambda_{12}vv'& 4\lambda_{13}uv'+\frac{f_2w}{2}&
-m_{\xi_\sigma}^2&
%8(2\lambda_{10} +\lambda_{11})v'^2-\frac{f_2uw}{2v'}&
4\lambda_{14}wv'+\frac{f_2u}{2}\\
 2\lambda_5vw+\frac{f_1u}{2}&
 2\lambda_6uw+\frac{f_1v}{2}+\frac{f_2v'}{2}&
4\lambda_{14}wv'+\frac{f_2u}{2}&
-m_{\xi_\chi}^2
% 4\lambda_3w^2-\frac{u}{2w}(f_1 v - f_2 v')
\end{array}\right),
\label{scalm}
\end{equation}
where $m_{\xi_\sigma}^2\equiv
-8(2\lambda_{10} +\lambda_{11})v'^2+\frac{f_2uw}{2v'},
m_{\xi_\chi}^2\equiv -4\lambda_3w^2+\frac{u}{2w}(f_1 v - f_2 v')$.

With this matrix, it is difficult to get a clear physical meaning.
As in~\cite{ton} a meaningful approximation is to impose $\vert f_1\vert,
\vert f_2\vert \sim w$  and to maintain only
terms of the second order in $w$ in Eq. (\ref{scalm})
(This means that we are working in low--energy phenomenology).
This procedure immediately gives us  one physical field
\begin{equation}
H_\chi \approx \xi_\chi
\label{h4}
\end{equation}
with mass
\begin{equation}
m^2_{H_\chi} \approx - 4 \lambda_3 w^2,
\label{mass3}
\end{equation}
and the square mass matrix of  $\xi_\eta, \xi_\rho, \xi_\sigma$ mixing
\begin{equation}
M^2_{3\xi} \approx \bordermatrix{& \xi_\eta & \xi_\rho & \xi_\sigma\cr
\xi_\eta&-\frac{f_1uw}{2v} &  \frac{f_1w}{2} & 0  \cr
\xi_\rho&\frac{f_1w}{2} &- \frac{w}{2u}(f_1v+f_2v')& \frac{f_2w}{2}\cr
\xi_\sigma&0& \frac{f_2w}{2}  &- \frac{f_2uw}{2v'}\cr}.
\label{scalmo}
\end{equation}
Solving the charateristic equation
of the matrix (\ref{scalmo})   we get one massless
field $H_1$ and two physical  ones ($H_2, H_3$) with masses
\begin{eqnarray}
x_{2,3}& =& -\frac{w}{4}\left[\frac{f_1}{vu}(v^2+u^2)+
\frac{f_2}{uv'}(v'^2+u^2)\right]\nonumber\\
&\pm&\frac{w}{4}\left\{\left[\frac{f_1}{vu}(v^2+u^2)+
\frac{f_2}{uv'}(v'^2+u^2)\right]^2 -
\frac{4f_1 f_2}{vv'}v^2_W\right\}^{1/2} \equiv m^2_{H_{2,3}}.
\label{sol23}
\end{eqnarray}
Here, in Figs. 1 and 2, we plot $m_{02} \equiv m_{H_2}/w$ and  
$m_{03} \equiv m_{H_3}/w$ as functions of $u,v$  
\begin{figure*}[thb]
%\vspace*{1cm}
\centerline{\epsfxsize=12cm\epsffile{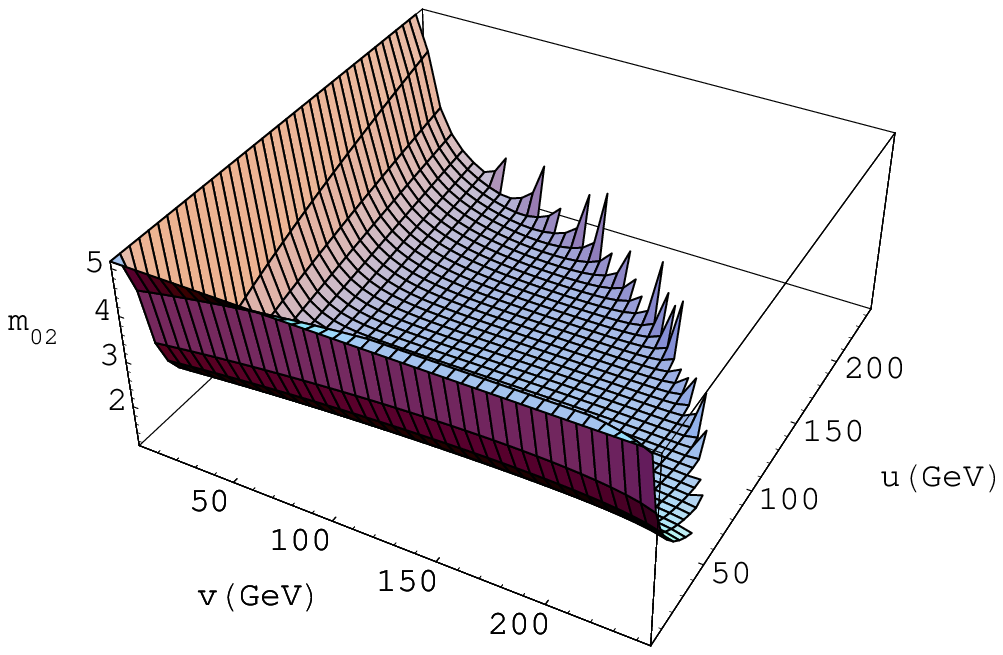}}
%\vspace*{1cm}
\caption{\label{fig1}{\em Behaviour of $m_{02}$ as a function of 
$u$ and $v$.}}
\vspace*{4mm}
%%\end{figure*}

%%\begin{figure*}[hbt]
\centerline{\epsfxsize=12cm\epsffile{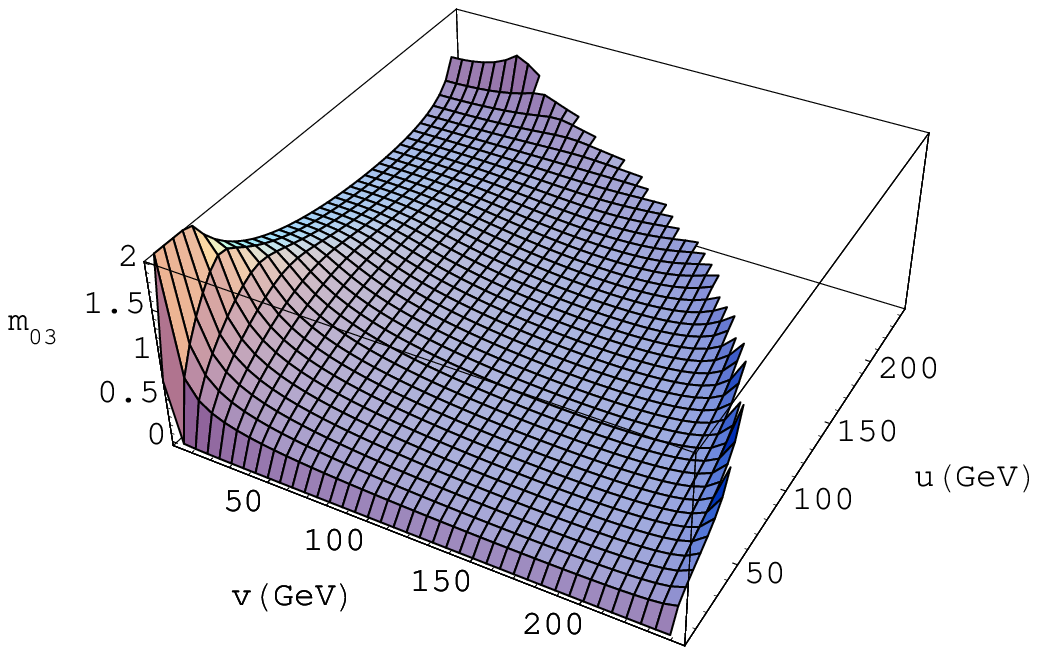}}
\caption{\label{fig2}{\em Behaviour of $m_{03}$ as a function of 
$u$ and $v$.}}
\end{figure*}
only since $v'$ is given according to the constrain 
$u^2 + v^2 + v'^2 = 246^2$. From the Figures, it 
follows that $u$ and $v$ are bounded as follows: 
$ 0 < u, v \leq 240$ GeV. Fig. 1 shows that $m_{H_2}$
increases when $u$ and $v$ tend to zero, while from Fig. 2 we see
that $m_{H_3}$ vanishes at $u=0$ and increases when $v$ tends to zero
and this mass is quite small. Note that our graphic surfaces are obtained
in the suggestion $f_1, f_2 \sim -w$ only. Otherwise, either 
$m_{03}$ is undefined (in the case of $f_1$ and $f_2$ having different signs) 
or both $m_{02}$ and $m_{03}$ are undefined (in case both $f_1$ and $f_2$ are 
positive). 
 
  The characteristic equation corresponding to $x_{2,3}$ can be written in
the compact form
\begin{equation}
u F_1(n) F_2(n) + v' F_1(n) + v F_2 (n) =0, ~~ n= 2,3,
\end{equation}
where
\begin{equation}
F_1(i)=\frac{2x_i}{f_1w} + \frac{u}{v} ~~\mbox{and}~~
F_2(i)=\frac{2x_i}{f_2w} + \frac{u}{v'}.
\end{equation}

To construct  physical fields (eigenstates) we begin from the
charateristic equation
\begin{equation}
\left( M^2  -  x_i \right) H_i = 0,\ i=1,2,3,
\label{char}
\end{equation}
where $M^2$ is the considered square mass matrix and $H_i \equiv
(H_{i1}, H_{i2}, H_{i3})^T$. With $M^2_{3\xi}$ we have a
system of three equations
%\begin{mathletters}
\begin{eqnarray}
-\left(\frac{f_1uw}{2v} + x_i\right) H_{i1} +
\frac{f_1w}{2} H_{i2} &=&0,
\label{root1}\\
\frac{f_1w}{2} H_{i1} -  \left[\frac{w}{2u}(f_1v+f_2v') +
x_i\right] H_{i2} + \frac{f_2w}{2} H_{i3}&=&0,
\label{root2}\\
\frac{f_2w}{2} H_{i2} - \left(\frac{f_2uw}{2v'} +
x_i\right) H_{i3}&=&0.
\label{root3}
\end{eqnarray}
%\end{mathletters}
It is known ~\cite{smir} that this system of equations is over defined
and must be reduced to two equations, in our case, the first and the
last. Let us drop the second equation, and suppose
\begin{equation}
H_{i1} =k(i),
\label{sup1}
\end{equation}
where $k(i)$ will be defined by the normalization 
of the states~\cite{smir}.\par
From Eq. (\ref{root1}) we obtain
\begin{equation}
 H_{i2} = \left( \frac{u}{v} + \frac{2 x_i}{f_1w}\right) k(i)
\equiv F_1(i)k(i),
\hspace*{2cm}  \mbox{(no summation over i)}.
\label{h2}
\end{equation}
Combining Eqs. (\ref{root3}) and (\ref{h2}) we then find
\begin{equation}
 H_{i3} = \frac{F_1(i)}{F_2(i)}k(i).
\label{h3}
\end{equation}
It is easy to see that from the condition of normalization of the
states $H_i$, ~ $k(i)$ is found  to be
\begin{equation}
 k(i) = \left [1 + F_1(i)^2 + \frac{F_1(i)^2}{F_2(i)^2}
\right ]^{-1/2}.
\label{nor1}
\end{equation}
Thus, we obtain finally a formula for the eigenstates of
$\xi_\eta, \xi_\rho, \xi_\sigma$ mixing
\begin{equation}
H_i = k(i) \left( \begin{array}{c}
1\\
F_1(i)\\
F_1(i)/F_2(i)
\end{array}
\right)\equiv
 \left( \begin{array}{c}
H_{i1}\\
H_{i2}\\
H_{i3}\\
\end{array}
\right).
\label{rootc}
\end{equation}
It is not difficult to verify that $H_i$ given in (\ref{rootc})
are orthogonal to each other.\par

  In the massless approximation  $i=1$ (i.e., $x_1=0$) we immediately find
\begin{equation}
H_1 = \frac{1}{\sqrt{v^2+u^2+v'^2}}
\left( \begin{array}{c}
v\\
u\\
v'\\
\end{array}
\right).
\label{rooto}
\end{equation}
In the next approximation (the $\lambda$'s are taken into account) 
the field $H_1$ acquires a mass. Following~\cite{ton} we
solve the characteristic  equation with the exact 3 $\times $ 3 mass matrix
$M^2_{3\xi}$, and the  $H_1$ associated with, namely
\begin{equation}
\left( M^2_{3\xi} - x_1 \right) H_1 = 0.
\label{slh1}
\end{equation}
From   system (\ref{slh1}) we obtain the following formulas for
the  $H_1$ mass
\begin{eqnarray}
m^2_{H_1}& =& x_1 \approx 4\lambda_1 v^2 + 2\lambda_4 u^2 +
4\lambda_{12} v'^2 \approx 2\lambda_4 v^2 + 4\lambda_2 u^2 +
4\lambda_{13} v'^2 \nonumber\\
&\approx & 4\lambda_{12} v^2 + 4\lambda_{13} u^2 +
8(2\lambda_{10} +\lambda_{11}) v'^2 .
\label{htkl}
\end{eqnarray}
From Eqs. (\ref{htkl}) we can accept the following relation
among coupling constants
\begin{equation}
\lambda \approx \lambda_1 \approx \lambda_{12} \approx
\lambda_4/2 \approx \lambda_2 \approx \lambda_{13} \approx
2(2\lambda_{10} + \lambda_{11}),
\label{cop}
\end{equation}
and then the mass of $H_1$
\begin{equation}
m^2_{H_1} \approx 4 \lambda (v^2 + u^2 +v'^2) \equiv M^2_{01}.
\label{mash1}
\end{equation}
Thus, according to (\ref{h4}), (\ref{rootc}) and (\ref{mash1}),
the eigenstates can be expressed as follows
%\begin{mathletters}
\begin{eqnarray}
\left(\begin{array}{cccc}
      H_1 \\
      H_2 \\
      H_3
\end{array}\right) & \approx &
\left(\begin{array}{cccc}
2\sqrt{\vert\lambda\vert}v/M_{01} &
2\sqrt{\vert\lambda\vert}u/M_{01} &
2\sqrt{\vert\lambda\vert}v^\prime/M_{01} \\
H_{21} & H_{22} & H_{23} \\
H_{31} & H_{32} & H_{33}
\end{array}\right)
\left(\begin{array}{cccc}
   \xi_\eta \\
   \xi_\rho \\
   \xi_\sigma
\end{array}\right),
\label{scalrt1} \\[2mm]
H_\chi  & \approx & \xi_\chi.
\label{scalrt2}
\end{eqnarray}
%\end{mathletters}
Since the matrix
\begin{equation}
A_{H\xi} =
\left(\begin{array}{cccc}
2\sqrt{\vert\lambda\vert}v/M_{01} &
2\sqrt{\vert\lambda\vert}u/M_{01} &
2\sqrt{\vert\lambda\vert}v^\prime/M_{01} \\
H_{21} & H_{22} & H_{23} \\
H_{31} & H_{32} & H_{33}
\end{array}\right)\equiv \left (A_{H\xi}^{-1}\right )^T ,
~~ \mbox{det}A_{H\xi}=1
\label{matrix}
\end{equation}
in (\ref{scalrt1}) is an orthogonal matrix $SO(3)$ the relation inverse
to (\ref{scalrt1}) and (\ref{scalrt2}) can easily be found~\cite{kor}
\begin{eqnarray}
\left(\begin{array}{cccc}
\xi_\eta \\
\xi_\rho \\
\xi_\sigma
\end{array}\right) & \approx &
\left(\begin{array}{cccc}
2\sqrt{\vert\lambda\vert}v/M_{01} &
H_{21} & H_{31} \\
2\sqrt{\vert\lambda\vert}u/M_{01}
 & H_{22} & H_{32} \\
2\sqrt{\vert\lambda\vert}v^\prime/M_{01}  & H_{23}  & H_{33}
\end{array}\right)
\left(\begin{array}{cccc}
H_1 \\
H_2 \\
H_3
\end{array}\right),
\label{scalrt3} \\[2mm]
\xi_\chi  & \approx & H_\chi .
\label{scalrt4}
\end{eqnarray}

   It is clear that our results here, especially the matrix in Eq.
(\ref{scalrt3}), are different from and more transparent than those
given by Eqs. (19a)--(23b) in~\cite{ton}. In addition to this, the
eigenstate corresponding to mass $-4\lambda_3 w^2$ is the scalar part of
$\chi^o$, while in~\cite{ton} it is the scalar part of $\sigma^o_2$.\par

  Similarly, in the pseudoscalar sector we obtain one physical
field $\zeta_\sigma \equiv \zeta'_\sigma$ with a mass equal
to the mass of $H'_\sigma$, and the square mass
matrix of the $\zeta_\eta, \zeta_\rho, \zeta_\sigma,  \zeta_\chi$ mixing
\begin{equation}
M^2_{4\zeta} = \bordermatrix{&\zeta_\eta& \zeta_\rho& \zeta_\sigma&
\zeta_\chi\cr
\zeta_\eta&-\frac{f_1uw}{2v}&-\frac{f_1w}{2} &0&-\frac{f_1u}{2} \cr
\zeta_\rho&-\frac{f_1w}{2}&-\frac{w}{2u}(f_1v+f_2v')&-\frac{f_2w}{2}&
 -\frac{1}{2}(f_1 v+f_2v') \cr
\zeta_\sigma&0& -\frac{f_2w}{2}&-\frac{f_2uw}{2v'}&-\frac{f_2u}{2}\cr
\zeta_\chi&-\frac{f_1u}{2}& -\frac{1}{2}(f_1 v+f_2v')&
-\frac{f_2u}{2}&-\frac{u}{2w}(f_1v+f_2v')\cr}.
\label{psem}
\end{equation}
With the  approximation  as mentioned above we obtain one Goldstone boson
 $G_1 \approx \zeta_\chi$ and the  $\zeta_\eta, \zeta_\rho,
\zeta_\sigma$ mixing
\begin{equation}
M^2_{3\zeta} = \bordermatrix{&\zeta_\eta & \zeta_\rho & \zeta_\sigma\cr
\zeta_\eta &-\frac{f_1uw}{2v} & - \frac{f_1w}{2} & 0\cr
\zeta_\rho&- \frac{f_1w}{2} &- \frac{w}{2u}(f_1v+f_2v')&- \frac{f_2w}{2}\cr
\zeta_\sigma&0&- \frac{f_2w}{2}  &- \frac{f_2uw}{2v'}\cr}.
\label{psemo}
\end{equation}
It can be checked that the characteristic equation in this
case have the same roots, but a different set of the eigenstates
(simply, make a replace $H_{i2} \rightarrow - H_{i2}$)
\begin{eqnarray}
\left(\begin{array}{cccc}
      A_1 \\
      A_2 \\
      A_3
\end{array}\right) & \approx &
\left(\begin{array}{cccc}
2\sqrt{\vert\lambda\vert}v/M_{01} &
-2\sqrt{\vert\lambda\vert}u/M_{01} &
2\sqrt{\vert\lambda\vert}v^\prime/M_{01} \\
 H_{21} &- H_{22} & H_{23} \\
H_{31} & -H_{32} & H_{33}
\end{array}\right)
\left(\begin{array}{cccc}
  \zeta_\eta \\
   \zeta_\rho \\
   \zeta_\sigma
\end{array}\right)
\label{psert1}
\end{eqnarray}
or equivalently
\begin{eqnarray}
\left(\begin{array}{cccc}
\zeta_\eta \\
\zeta_\rho \\
\zeta_\sigma
\end{array}\right) & \approx &
\left(\begin{array}{cccc}
2\sqrt{\vert\lambda\vert}v/M_{01} &
  H_{21} & H_{31}\\
-2\sqrt{\vert\lambda\vert}u/M_{01} &- H_{22} & -H_{32} \\
2\sqrt{\vert\lambda\vert}v^\prime/M_{01}  & H_{23} & H_{33}
\end{array}\right)
\left(\begin{array}{cccc}
 A_1 \\
 A_2 \\
 A_3
\end{array}\right).
\label{psert2}
\end{eqnarray}

  In the singly charged sector the mixing occurs in the set of
$ \eta^+_1, \rho^+, s^+_1$  and in the set of $ \eta^+_2, \chi^+, s^+_2$
(while in~\cite{ton} the decompositions are $ \eta^+_1, \rho^+,\eta^+_2$
and $\chi^+, s^+_1, s^+_2$) with the following square mass matrices
\begin{equation}
M^2_{+1} = \bordermatrix{&\eta^+_1 & \rho^+& s^+_1&\cr
\eta^-_1&\lambda_7u^2-\frac{f_1uw}{2v}&\lambda_7vu-\frac{f_1w}{2}&0 \cr
\rho^-&\lambda_7vu-\frac{f_1w}{2}&\lambda_7v^2-\frac{w}{2u}(f_1v+f_2v')
&\frac{f_2w}{2}\cr
s^-_1&0&\frac{f_2w}{2}&   -\frac{f_2uw}{2v'}\cr},
\label{sinm1}
\end{equation}
and
\begin{equation}
M^2_{+2} = \bordermatrix{& \eta^+_2& \chi^+&  s^+_2& \cr
\eta^-_2&\lambda_8w^2-\frac{f_1uw}{2v}&\lambda_8vw-\frac{f_1u}{2}&0\cr
\chi^-&\lambda_8vw-\frac{f_1u}{2}&\lambda_8v^2-\frac{u}{2w}(f_1v+f_2v')&
\frac{f_2u}{2}\cr
s^-_2&0&\frac{f_2u}{2}&-\frac{f_2uw}{2v'}\cr}.
\label{sinm2}
\end{equation}
Applying the above approximation to $M^2_{+2}$ we obtain one Goldstone boson
$G^+_2 \approx \chi^+$ and two physical fields associated with
$\eta^+_2$ and $s^+_2$ and masses
\begin{equation}
m^2_{\eta^+_2} = -\lambda_8w^2+\frac{f_1uw}{2v},\hspace*{0.2cm}
m^2_{s^+_2} = \frac{f_2uw}{2v'}.
\label{mass5}
\end{equation}
For the   $\eta^+_1, \rho^+, s^+_1$ mixing, we have
\begin{equation}
M^2_{+1} = \bordermatrix{& \eta^+_1& \rho^+& s^+_1\cr
\eta^-_1&-\frac{f_1uw}{2v} & - \frac{f_1w}{2} & 0  \cr
\rho^-&- \frac{f_1w}{2} &- \frac{w}{2u}(f_1v+f_2v')& \frac{f_2w}{2}\cr
s^-_1&0& \frac{f_2w}{2}  &- \frac{f_2uw}{2v'}\cr}.
\label{sinmo}
\end{equation}
As before, the characteristic equation of (\ref{sinmo}) has the
same roots, but the eigenstates are different and are  given by
(necessary replaces: $H_{i2} \rightarrow - H_{i2},  H_{i3}
\rightarrow - H_{i3}$)
\begin{eqnarray}
\left(\begin{array}{cccc}
      h^+_1 \\
      h^+_2 \\
      h^+_3
\end{array}\right) & \approx &
\left(\begin{array}{cccc}
2\sqrt{\vert\lambda\vert}v/M_{01} &
-2\sqrt{\vert\lambda\vert}u/M_{01} &
-2\sqrt{\vert\lambda\vert}v^\prime/M_{01} \\
H_{21} &- H_{22} &- H_{23} \\
 H_{31} &- H_{32} &- H_{33}
\end{array}\right)
\left(\begin{array}{cccc}
   \eta^+ \\
   \rho^+ \\
   s^+_1
\end{array}\right)
\label{sinrt1}
\end{eqnarray}
or equivalently
\begin{eqnarray}
\left(\begin{array}{cccc}
\eta^+ \\
\rho^+ \\
s^+_1
\end{array}\right) & \approx &
\left(\begin{array}{cccc}
2\sqrt{\vert\lambda\vert}v/M_{01} &
H_{21} &  H_{31} \\
-2\sqrt{\vert\lambda\vert}u/M_{01}  &- H_{22} &- H_{32} \\
-2\sqrt{\vert\lambda\vert}v^\prime/M_{01}  &- H_{23} &- H_{33}
\end{array}\right)
\left(\begin{array}{cccc}
 h^+_1 \\
 h^+_2 \\
 h^+_3
\end{array}\right).
\label{sinrt2}
\end{eqnarray}

  In the doubly charged sector the mixing occurs up
all states   $\rho^{++},s^{++}_2, \chi^{++}, s^{++}_1$,
and the square mass matrix is given
\begin{equation}
M^2_{4++} = \left(
\begin{array}{cccc}
\lambda_9 w^2-\frac{w}{2u}(f_1v+f_2v')&\frac{f_2w}{2}
&\lambda_9uw-\frac{f_1v}{2}+\frac{f_2v'}{2} &0 \\
\frac{f_2w}{2}&2\lambda_{11}v'^2-\frac{f_2uw}{4v'}
&0&2\lambda_{11}v'^2\\
\lambda_9uw-\frac{f_1v}{2}+\frac{f_2v'}{2}&0&
m^2_{\chi^{++} \chi^{--}}&
\frac{f_2u}{2}\\
0&2\lambda_{11}v'^2& \frac{f_2u}{2}&
m^2_{s^{++}s^{--}}
\end{array}\right),
\label{dubm}
\end{equation}
where $m^2_{\chi^{++} \chi^{--}}\equiv \lambda_9
u^2-\frac{u}{2w}(f_1v+f_2v')$ and $ m^2_{s^{++}s^{--}}\equiv 2
\lambda_{11}v'^2-\frac{f_2uw}{4v'}$.\par
By the same way as mentioned above
we obtain one Goldstone boson $G_3^{++}
\approx \chi^{++}$ and one physical field $s^{++}_1$ with mass
\begin{equation}
m^2_{s^{++}_1} = \frac{f_2uw}{4v'},
\label{mass8}
\end{equation}
and a matrix of $\rho^{++}, s_2^{++}$ mixing
\begin{equation}
M^2_{2++} = \bordermatrix{& \rho^{++}& s_2^{++}\cr
\rho^{--}&\lambda_9w^2 -  \frac{w}{2u}(f_1v+f_2v')&\frac{f_2w}{2}  \cr
s_2^{--}& \frac{f_2w}{2}&- \frac{f_2uw}{4v'}\cr}.
\label{dubmo}
\end{equation}
Solving the characteristic equation we get two physical fields
\begin{equation}
\left(\begin{array}{c}
d_1^{++}\\
d_2^{++}\\
\end{array}\right) =
\left( \begin{array}{cc}
 n_4\left(\frac{u}{2v'} + \frac{2x_4}{2f_2w}\right),  & n_4\\
 n_5\left(\frac{u}{2v'} + \frac{2x_5}{2f_2w}\right),  & n_5
\end{array} \right)
\left(\begin{array}{c}
\rho{++} \\
s_2^{++} \end{array} \right),
\end{equation}
 where
\begin{equation}
 n(i) = \left[1 +  \left( \frac{u}{2v'} + \frac{2 x_i}{f_2w}\right)^2
\right]^{-\frac{1}{2}} (i=4,5),
\label{nor2}
\end{equation}
with masses
\begin{eqnarray}
x_{4,5}& =&\frac{w}{2}\left[
\lambda_9w-\frac{f_2}{2}\left(\frac{u}{2v'}+
\frac{v'}{u}\right)-\frac{f_1v}{2u}\right]\nonumber\\
&\pm&\frac{w}{2}\left\{\left[
\lambda_9w-\frac{f_2}{2}\left(\frac{u}{2v'}+
\frac{v'}{u}\right)-\frac{f_1v}{2u}\right]^2+\frac{\lambda_9 f_2u w}{v'}
+\frac{f_2}{2v'}(f_2v'-f_1v)\right\}^{1/2}.
\label{sol67}
\end{eqnarray}
\vspace*{2mm}

Let us summarize the particle content in the considered  Higgs sector:\\

-- in the neutral scalar sector  physical fields are: $H_1, H_2, H_3,
H'_\sigma$ and $H_\chi$\
\begin{eqnarray}
m^2_{H_1}& \approx& - 4\lambda v^2_W,\ m^2_{H_2}= x_2,\
m^2_{H_3} = x_3, \nonumber\\
  m^2_{H'_\sigma}&=& 2\lambda_{11}v'^2 +
\frac{f_2 u w}{4v'},\
 m^2_\chi \approx - 4\lambda_3 w^2,
\label{m1}
\end{eqnarray}

-- in the neutral pseudoscalar sector, physical fields are:
 $A_2, A_3, A_\sigma$ and two Goldstone bosons:
$G_1 \approx \zeta_\chi$ and $ G_2$ corresponding to the  massless $A_1$
\begin{equation}
 m^2_{A_2}= x_2,\
m^2_{A_3} = x_3, \
  m^2_{A'_\sigma}= m^2_{H'_\sigma},
\label{m2}
\end{equation}

-- in singly charged sector there are two Goldstone bosons $G_3 = h^+_1,
G^+_4 \approx \chi^+$ and three physical fields :
$h^+_2, h^+_3, \eta^+_2, s^+_2$ with masses:
\begin{equation}
m^2_{h^+_2}=m^2_{H_2},\  m^2_{h^+_3}=m^2_{H_3},\
m^2_{\eta^+_2} = -\lambda_8w^2+\frac{f_1uw}{2v},\hspace*{0.2cm}
m^2_{s^+_2} = \frac{f_2uw}{2v'},
\label{m3}
\end{equation}

-- in doubly charged sector we have one Goldstone ($G_5^{++}
\approx \chi^{++}$) and three physical fields  $ d_1^{++}, d_2^{++},
s_1^{++}$ with masses:
\begin{equation}
m^2_{d^{++}_1} = x_4, m^2_{d^{++}_2} = x_5,
m^2_{s^{++}_1} = \frac{f_2uw}{4v'}.
\label{m4}
\end{equation}
Eqs. (\ref{m1} -- \ref{m4}) show that $f_2$ has to be positive and there are
 three degenerate states $H_2, A_2$ and $h^+_2$ in mass $x_1$,
another three degenerate states $H_3, A_3$ and $h^+_3$ in mass $x_2$, and
two degenerate ones $H'_\sigma, A'_\sigma$ in mass $m^2_{H'_\sigma}$.
Again, eigenstates in these sectors (singly and doubly charged)
are different from those in~\cite{ton}\par
Combining conditions for consistency (\ref{cop}) and positiveness
of the mass square the following bounds for coupling constants
are followed
%\begin{mathletters}
\begin{eqnarray}
\lambda \approx \lambda_1 \approx \lambda_{12} \approx
\lambda_4/2& \approx &\lambda_2 \approx \lambda_{13} \approx
2(2\lambda_{10} + \lambda_{11})  \stackrel{>}{\sim} 0 ,
\label{dif}\\
\lambda_3& \stackrel{<}{\sim}& 0.
\end{eqnarray}
%\end{mathletters}
It is worth mentioning here that the last relation
$\lambda \approx 2(2\lambda_{10} + \lambda_{11})$ was
replaced by $\lambda \approx 4(2\lambda_{10} + \lambda_{11})$
in~\cite{ton} (see Eq. (21) there).

\section{Conclusion}

\hspace*{0.5cm}  In conclusion, we have considered  in this paper 
the Higgs sector of the
minimal 3 3 1 model in the  condition $|f_1|, |f_2| \sim w \gg v, u, v'$.
However, a consistent sign of $f_2$ is still under
question: a positive $m^2_{H_3}$  needs negative $f_1$ and $f_2$ 
(see Eq. (\ref{sol23})), while expressions in Eqs. (\ref{mass5}) -- (\ref{m4}) 
require $f_2$ to be positive. This problem deserves further studies. 

%which indicates that trilinear Higgs boson couplings are much larger
%than quartic Higgs boson couplings. 
Other possibilities such as
$f_1, f_2 \sim v,u$ or $v'$ cannot give us a simple solution.
We do hope that further studies will justify this assumption.\par

  It is to be mentioned that exposed in a more transparent way our results
(eigenvalues and eigenstates) in the model of three triplets and one
sextet, have some differences from those of the 
author of ~\cite{ton}. Our graphic surfaces give the conclusions, for example, 
$0 < u,v \leq 240$ GeV, quite different from those in~\cite{ton}.
However, there is a mass degeneracy in mass of  scalar, pseudoscalar  
and singly charged Higgs fields with mass $m_{H_3}$ relatively small. 

In the toy model -- the three triplet model,  the
unnatural condition $v\neq u$ was removed. We hope that these differences
will be examined in the future.

{\it Note added}~: After submitting this paper for publication we have
just been informed that some of our conclusions are in agreement with 
results in Ref.\cite{gum}.\par 
{\bf Acknowledgments}

The authors thank D. Gomez Dumm for bringing their attention to 
Ref.~\cite{gum} and for a discussion on some of the results. 
  H. N. L. thanks T. Inami and Department of Physics, Chuo 
University for warm hospitality. 
N.A.K. thanks CNRS for financial support under the joint France--Vietnam
Convention Internationale. He also thanks LAPTH for warm hospitality and 
P. Aurenche and G. Belanger for discussions.\par 
This work was supported in part by  the Research Programme on Natural 
Sciences of Hanoi National University under grant number QT 98.04 and 
KT - 04.1.1.

%\section*{Appendix A}
\appendix
\renewcommand{\thesection}{Appendix {\Alph{section}}}
\section{}
\renewcommand{\thesection}{{\Alph{section}}}
\makeatletter
\setcounter{secnumdepth}{5}
\setcounter{tocdepth}{5}
\renewcommand{\theequation}{\thesection.\arabic{equation}}
\@addtoreset{equation}{section}
\makeatother

Here we give the full expression for the potential
\begin{eqnarray}
&&V_S\left(\eta ,\rho ,\chi,S\right)=
\mu_1^2(v^2+2v\xi_\eta+\xi^2_\eta+\zeta^2_\eta+\eta^+_1\eta^-_1+
\eta^+_2\eta^-_2)\nonumber\\
&&+\mu_2^2(u^2+2u\xi_\rho+\xi^2_\rho+\zeta^2_\rho+\rho^+\rho^-+
\rho^{++}\rho^{--})\nonumber\\
&&+\mu_3^2(w^2+2w\xi_\chi+\xi^2_\chi+\zeta^2_\chi+\chi^+\chi^-+
\chi^{++}\chi^{--})\nonumber\\
&&+\lambda_1(v^2+2v\xi_\eta+\xi^2_\eta+\zeta^2_\eta+\eta^+_1\eta^-_1+
\eta^+_2\eta^-_2)^2
+\lambda_2(u^2+2u\xi_\rho+\xi^2_\rho+\zeta^2_\rho+\rho^+\rho^-+
\rho^{++}\rho^{--})^2\nonumber\\
&&+\lambda_3(w^2+2w\xi_\chi+\xi^2_\chi+\zeta^2_\chi+\chi^+\chi^-+
\chi^{++}\chi^{--})^2\nonumber\\
&&+(v^2+2v\xi_\eta+\xi^2_\eta+\zeta^2_\eta+\eta^+_1\eta^-_1+
\eta^+_2\eta^-_2)\nonumber\\
&&\times[\lambda_4(u^2+2u\xi_\rho+\xi^2_\rho+\zeta^2_\rho+\rho^+
\rho^-+\rho^{++}\rho^{--})\nonumber\\
&&+\lambda_5 (w^2+2w\xi_\chi+\xi^2_\chi+\zeta^2_\chi+\chi^+\chi^-+
\chi^{++}\chi^{--})]\nonumber\\
&&+\lambda_6(u^2+2u\xi_\rho+\xi^2_\rho+\zeta^2_\rho+\rho^+\rho^-+
\rho^{++}\rho^{--})
(w^2+2w\xi_\chi+\xi^2_\chi+\zeta^2_\chi+\chi^+\chi^-+
\chi^{++}\chi^{--})\nonumber\\
&&+\lambda_7(\eta^{o2}\rho^+\rho^-+\eta^o\rho^o
\eta^+_1\rho^-+\eta^o\rho^-
\eta^-_2\rho^{++}+\eta^{o*}\rho^{o*}\rho^+\eta^-_1+
\rho^{o2}\eta^+\eta^-\nonumber\\
&&+\rho^{o*}\eta^-_1\eta^-_2\rho^{++} + \eta^{o*}\rho^+
\eta_2^+\rho^{--} + \rho^o\eta^+_1\eta^+_2\rho^{--}+
\eta_2^+\eta_2^-\rho^{++}\rho^{--})\nonumber\\
&&+\lambda_8(\eta^{o2}\chi^+\chi^- + \chi^{o2}\eta^+_2\eta^-_2
+ \eta^o\chi^o\eta^-_2 \chi^+ + \eta^{o*}\chi^{o*}
\eta_2^+\chi^- + \eta^o\eta^+_1\chi^+\chi^{--} \nonumber\\
&&+\eta^{o*}\eta^-_1\chi^-\chi^{++} + \chi^{o*}\eta_1^+
\eta_2^+\chi^{--} + \chi^o\eta^-_1\eta^-_2\chi^{++} +
\eta_1^+\eta_1^-\chi^{++}\chi^{--})\nonumber\\
&&+\lambda_9(\chi^{o2}\rho^{++}\rho^{--}+\rho^{o2}
\chi^{++}\chi^{--}+\rho^o\chi^o\chi^{++}\rho^{--}+
\rho^{o*}\chi^{o*}\rho^{++}\chi^{--}\nonumber\\
&&+\rho^o\rho^-\chi^-\chi^{++}+\rho^{o*}\rho^+
\chi^+\chi^{--}+\chi^o\rho^+\chi^+\rho^{--}+\chi^{o*}
\rho^-\chi^-\rho^{++}+\rho^+\rho^-\chi^+\chi^-)\nonumber\\
&&+\left\{\frac{f_1}{2}\left[(v+\xi_\eta+i\zeta_\eta)(\rho^o\chi^o -
\rho^{++}\chi^{--}) - \eta^-_1 (\chi^o\rho^+-\rho^{++}\chi^{--})
+\eta_2^+(\rho^+\chi^{--} -
\rho^o\chi^-)\right] + \mbox{h.c}\right\}\nonumber\\
&&+\mu_4^2[2(v'^2+2v'\xi_\sigma+\xi^2_\sigma+
\zeta^2_\sigma)+\xi'^2_\sigma+\zeta'^2_\sigma+2s_1^+s_1^- +
2s_2^+s_2^- + s_1^{++}s_1^{--} +  s_2^{++}s_2^{--} ] \nonumber\\
&&+\lambda_{10}[2(v'^2+2v'\xi_\sigma+\xi^2_\sigma+\zeta^2_\sigma)+
\xi'^2_\sigma+\zeta'^2_\sigma+2s_1^+s_1^- +
2s_2^+s_2^- + s_1^{++}s_1^{--} +  s_2^{++}s_2^{--} ]^2 \nonumber\\
&&+[2(v'^2+2v'\xi_\sigma+\xi^2_\sigma+\zeta^2_\sigma)+
\xi'^2_\sigma+\zeta'^2_\sigma+2s_1^+s_1^- +
2s_2^+s_2^- + s_1^{++}s_1^{--} +  s_2^{++}s_2^{--} ] \nonumber\\
&&\times[\lambda_{12}(v^2+2v\xi_\eta+\xi^2_\eta+\zeta^2_\eta+
\eta^+_1\eta^-_1+ \eta^+_2\eta^-_2)\nonumber\\
&&+\lambda_{13}(u^2+2u\xi_\rho+\xi^2_\rho+\zeta^2_\rho+\rho^+
\rho^-+\rho^{++}\rho^{--})\nonumber\\
&&+\lambda_{14} (w^2+2w\xi_\chi+\xi^2_\chi+\zeta^2_\chi+\chi^+\chi^-
+\chi^{++}\chi^{--})]\nonumber\\
&&+\left[\frac{f_2}{2}(\chi^-(\sigma^o_1\rho^++\rho^o s_2^+
+\rho^{++}s_1^-)+\chi^{--}(\rho^+s_2^++\rho^o s_1^{++}
+\sigma_2^o\rho^{++})\right.\nonumber\\
&&+\left. \chi^o(\sigma_2^o\rho^o
+\rho^+s_1^-+\rho^{++}s_2^{--})+\mbox{h.c}\right]\nonumber\\
&&+\lambda_{11}\left\{(\xi'^2_\sigma+\zeta'^2_\sigma +
s_1^+s_1^-+s_2^+s_2^-)^2+
(v'^2+2 v'\xi_\sigma + \xi^2_\sigma+\zeta^2_\sigma
+ s_2^+s_2^-+s_1^{++}s_1^{--})^2\right.\nonumber\\
&&+(v'^2+2v'\xi_\sigma + \xi^2_\sigma+\zeta^2_\sigma +
s_1^+s_1^-+s_2^{++}s_2^{--})^2\nonumber\\
&&+2[(v'^2+2 v'\xi_\sigma + \xi^2_\sigma+\zeta^2_\sigma)
s_1^+s_1^-+(\xi'^2_\sigma+\zeta'^2_\sigma)s_2^+s_2^-
+\sigma_1^o\sigma_2^os_1^+s_2^-+
\sigma_1^{o*}\sigma_2^{o*}s_2^+s_1^-\nonumber\\
&& +\sigma_1^{o*}s_2^+s_2^+s_1^{--}
+\sigma_1^{o}s_2^-s_2^-s_1^{++}
+\sigma_2^{o*}s_1^-s_2^-s_1^{++}+
\sigma_2^os_1^+s_2^+s_1^{--}+
s_2^+s_2^-s_1^{++}s_1^{--}]\nonumber\\
&&+2[(v'^2+2v'\xi_\sigma + \xi^2_\sigma+\zeta^2_\sigma)
s_2^+s_2^-+(\xi'^2_\sigma+\zeta'^2_\sigma)s_1^+s_1^-
+\sigma_1^o\sigma_2^os_1^+s_2^-+
\sigma_1^{o*}\sigma_2^{o*}s_2^+s_1^-\nonumber\\
&&+\sigma_1^{o^*}s_1^-s_1^-s_2^{++}
+\sigma_1^{o}s_1^+s_1^+s_2^{--}
+\sigma_2^{o*}s_1^+s_2^+s_2^{--}+
\sigma_2^os_1^-s_2^-s_2^{++}+
s_1^+s_1^-s_2^{++}s_2^{--}] \nonumber\\
&&+2[(v'^2+2 v'\xi_\sigma + \xi^2_\sigma+\zeta^2_\sigma)
(s_1^{++}s_1^{--}+s_2^{++}s_2^{--})
+\sigma_2^o\sigma_2^os_2^{++}s_1^{--}+
\sigma_2^{o*}\sigma_2^{o*}s_1^{++}s_2^{--}\nonumber\\
&&+\sigma_2^os_1^-s_2^-s_2^{++}+
\sigma^o_2s^+_1s^+_2s^{--}_1
+\left.\sigma_2^{o*}s_1^+s_2^+s_2^{--}
+\sigma_2^{o*}s_1^-s_2^-s_1^{++}
+s_1^+s_1^-s_2^{+}s_2^{-}]\right\}.
\label{A.1}
\end{eqnarray}
%in which $\xi_1, \xi_2, \xi_3, \xi_4, H_5$ stand for
%$\xi_\eta, \xi_\rho, \xi_\chi, \xi_\sigma, \xi'_\sigma$
%respectively.

%\section*{Appendix B}

\renewcommand{\thesection}{Appendix {\Alph{section}}}
\section{}
\renewcommand{\thesection}{{\Alph{section}}}
\makeatletter
\setcounter{secnumdepth}{5}
\setcounter{tocdepth}{5}
\renewcommand{\theequation}{\thesection.\arabic{equation}}
\@addtoreset{equation}{section}
\makeatother

 The  constraint equations in the three triplet case (\ref{hig1})
with the potential $V_T$ are
\begin{eqnarray}
\mu^2_1 + 2 \lambda_1 v^2 + \lambda_4 u^2 + \lambda_5 w^2
+ \frac{f_1 u w}{2 v}&  = & 0, \nonumber\\
\mu^2_2 + 2 \lambda_2 u^2 + \lambda_4 v^2 + \lambda_6 w^2
+ \frac{f_1 v w}{2 u} &  = & 0,\label{B.1}\\
\mu^2_3 + 2 \lambda_3  w^2 + \lambda_5 v^2 + \lambda_6 u^2
+ \frac{f_1 v u }{2 w}&  = & 0\nonumber .
\end{eqnarray}
In the $\xi_\eta, \xi_\rho, \xi_\chi$ basis
(which is ortho--normalized) the square mass matrix,
after imposing the constraints (\ref{B.1}), has
the following form
\begin{equation}
M^2_{\xi} = \bordermatrix{& \xi_\eta & \xi_\rho & \xi_\chi\cr
\xi_\eta& 8 \lambda_1 v^2 - f_1 u w/v & 4 \lambda_4 v u + f_1 w &
 4 \lambda_5 v w + f_1 u \cr
\xi_\rho& 4 \lambda_4 v u + f_1 w &  8 \lambda_2 u^2 - f_1 v w/u&
 4 \lambda_6 u w + f_1 v \cr
\xi_\chi& 4 \lambda_5 v w + f_1 u & 4 \lambda_6 u w + f_1 v &
 8 \lambda_3 w^2 - f_1  v u / w \cr}
\left( \frac{1}{2}\right).
\label{B.2}
\end{equation}
Comparing with the results in~\cite{ton} we notice
that the mass matrix $M^2_{\xi}$ almost coincides with
the mass matrix given by Eq. (4) in~\cite{ton}.\par

 We will use the following approximation
\begin{equation}
|f_1| \sim w, \hspace*{1cm} w \gg v, u.
\label{B.3}
\end{equation}
Using (\ref{B.3}) and keeping only the terms of second order in $w$,
we got one  massless ($H_1$) and two massive physical
states ($H_2$ and $H_3$) with masses
\begin{equation}
 m^2_{H_2} \approx \frac{v^2 + u^2}{2 v u} w^2
~~ \mbox{and} ~~
m^2_{H_3} \approx - 4 \lambda_3 w^2,
\label{B.4}
\end{equation}
and mixing
\[
\left(\begin{array}{c}
\xi_\eta\\
\xi_\rho\\
\end{array}\right) \approx \frac{1}{(v^2 + u^2)^{1/2}}
\left( \begin{array}{cc}
v  & - u\\
 u & v\end{array} \right)
\left(\begin{array}{c}
H_1 \\
H_2 \end{array} \right),
\]
%\begin{equation}
\[\xi_\chi \approx H_3.\]
%\end{equation}
In order to improve the approximation, following~\cite{ton}
we search the mass of $H_1$ by solving the characteristic  equation
with the exact $3\times  3$ mass matrix $M^2_{\xi}$, and the  $H_1$
associated with
\begin{equation}
\left( M^2_{\xi} - m^2_{H_1} \right) H_1 = 0.
\label{B.5}
\end{equation}
Solving a system of three equations (\ref{B.5}) we get the mass for $H_1$
\begin{equation}
m^2_{H_1} \approx 4 \frac{\lambda_2 u^4 - \lambda_1 v^4}{v^2 - u^2},
\label{B.6}
\end{equation}
and relations among the coupling constants and VEVs
(if the approximation $f_1\approx -w$ is taken)
\begin{equation}
\lambda_4 \approx  2 \frac{\lambda_2 u^2 - \lambda_1 v^2}{v^2 - u^2},
\  \lambda_5 v^2 + \lambda_6 u^2 \approx - \frac{f_1 v u}{2 w} =
 \frac{v u}{2}.
\label{B.7}
\end{equation}

 Repeating the procedure for $H_2$ in a similar way we obtain an improved
approximation for the mass $m^2_{H_2}$
\begin{equation}
m^2_{H_2} \approx 4 v^2 u^2\frac{\lambda_1  - \lambda_2}{v^2 - u^2}
+f_1 w\frac{u^2 + v^2}{uv},
\label{B.8}
\end{equation}
where the second term in (B.8) is the zero--approximation (B.4). Besides
that we also obtain $\lambda_4$ given again as in (B.7) and
\begin{equation}
\lambda_5 -  \lambda_6  \approx  f_1 \frac{v^2-u^2}{4vuw} \approx
\frac{u^2-v^2}{4vu}.
\label{B.9}
\end{equation}

  We can solve (\ref{B.7}) and (\ref{B.9}) for $\lambda_5$ and $\lambda_6$

\begin{equation}
\lambda_5 = -\frac{f_1 u}{4vw} \approx \frac{u}{4v}, ~~
\lambda_6 = -\frac{f_1 v}{4uw}  \approx \frac{v}{4u}, ~~
\lambda_5. \lambda_6 = \left (\frac{f_1}{4w}\right )^2 \approx \frac{1}{16}.
\label{B.10}
\end{equation}

%% Using the latter formula we get from (\ref{B.6}) and (\ref{B.8})
%%\begin{equation}
%%m^2_{H_1} \approx 4 \frac{\lambda_2 u^4 - \lambda_1 v^4}{v^2 - u^2}
%%=,
%%\label{B.6'}
%%\end{equation}
%%and
%%\begin{equation}
%%m^2_{H_2} \approx 4 v^2 u^2\frac{\lambda_1  - \lambda_2}{v^2 - u^2}
%%+f_1 w\frac{u^2 + v^2}{uv},
%%\label{B.8'}
%%\end{equation}
%%respectively.\\

From (\ref{B.6}) and (\ref{B.8}) we see that if $v = u$ and $\lambda_1
\neq \lambda_2$, the expressions of $m^2_{H_1}$ and $m^2_{H_2}$ become
uncertain. It means
that $v = u$ is a special case, and let us consider this one.  \par
Setting $v = u$ and keeping only terms of second order in $w$,
we got again one  massless ($\hat{H}_1^o$) and two massive physical
states ($\hat{H}_2$ and $\hat{H}_3$) with masses
\begin{eqnarray}
\hat{H}_1 & \approx & \frac{1}{\sqrt{2}} (\xi_\eta + \xi_\rho);\
m^2_{\hat{H}_1} = 0,
\label{B.11}\\
\hat{H}_2 & \approx & \frac{1}{\sqrt{2}} (- \xi_\eta + \xi_\rho);\
m^2_{\hat{H}_2} = w^2,
\label{B.12}\\
\hat{H}_3 & \approx & \xi_\chi; \  m^2_{\hat{H}_3} = -
4\lambda_3 w^2.
\label{B.13}
\end{eqnarray}
From above equations,  we see that eigenstates in this case are
independent of $u$, and mass of  $\hat{H}_2$  depend only
on $w$ -- VEV of the heavy Higgs field  $\chi$  at the first
step of symmetry breaking.\par
In order to improve the approximation, as before,  we search mass of
$\hat{H}_1$ by solving the following equation
\begin{equation}
\left( M^2_{\xi}|_{v=u} -  m^2_{\hat{H}_1} \right) \hat{H}_1 = 0.
\label{B.14}
\end{equation}
From Eq. (\ref{B.14})  we get directly the following relations
\begin{eqnarray}
\lambda_1 & = & \lambda_2,\\
m^2_{\hat{H}_1} & = & 2 ( 2 \lambda_1 + \lambda_4 ) u^2,
\label{B.16}\\
\lambda_5 + \lambda_6 & = & -\frac{f_1}{2 w} \approx \frac{1}{2}.
\label{B.17}
\end{eqnarray}

  Now we consider  the pseudoscalar  sector.
In the $\zeta_\eta, \zeta_\rho, \zeta_\chi$ basis (which is also
ortho--normalized) $M^2_{\zeta}$ takes the form
\begin{equation}
M^2_{\zeta} = \bordermatrix{&\zeta_\eta & \zeta_\rho & \zeta_\chi\cr
\zeta_\eta&  u /v & 1 & u/w \cr
\zeta_\rho& 1 &  v/u& v/w\cr
\zeta_\chi&u/w & v/w & v u /w^2 \cr}(-f_1 w).
\label{B.18}
\end{equation}
It is easy  to check that, the diagonalization of 
 $M^2_{\zeta}$ gives us
two  Goldstone bosons ($G_2, G_3$)
and one massive pseudoscalar $A_1$ with exact mass
\begin{equation}
m^2_{A_1} = - \frac{f_1 w}{v u} \left[ v^2 + u^2 +
\left( \frac{v u}{w}\right)^2\right].
\end{equation}

  Let's denote the (normalized) eigenstate vectros $G_1$,
$G_2$ and $G_3$ with their coordinates $X_i, Y_i, Z_i$, ~ $i =  G_1, G_2,
G_3$, as follows
\begin{equation}
\left(\begin{array}{c}
X_i\\
Y_i\\
Z_i
\end{array}\right)\equiv
X_i \zeta_{\eta}+Y_i \zeta_{\rho}+Z_i \zeta_{\chi}, ~~
X_i^2 + Y_i^2 + Z_i^2 =1.
\label{B.20}
\end{equation}
%%where
%%$N_i=1/\sqrt{x_i^2+y_i^2+z_i^2}$, ~~
%% $N_i=(x_i^2+y_i^2+z_i^2)^{-1/2}$,  ~~
%%$N_i=(X_i^2 + Y_i^2 + Z_i^2)^{-1/2}$
%%are the normalization coefficients.
The characteristic equations of $M^2_{\zeta}$ written in the forms
\begin{equation}
\frac{X_n}{v} + \frac{Y_n}{u} + \frac{Z_n}{w} = 0, ~~ n=G_1, G_2
\label{B.21}
\end{equation}
\begin{equation}
vX_{G_3} = uY_{G_3} = wZ_{G_3}
\label{B.22}
\end{equation}
show that the massless
states $G_1$ and $G_2$ (orthogonal to each other, of course) belong
to a plane orthogonal to the vector $(1/v, 1/u, 1/w)^T$
%\begin{equation}
%\left(\begin{array}{c}
%1/v\\
%1/u\\
%1/w
%\end{array}\right)
%\end{equation}
%proved to be
which in turn is parallel to the massive state $G_3$
\begin{equation}
G_3 = \left(\begin{array}{c}
X_{G_3}\\
Y_{G_3}\\
Z_{G_3}
\end{array}\right)=N_{G_3}
\left(\begin{array}{c}
1/v\\
1/u\\
1/w\\
\end{array}\right),
\label{B.23}
\end{equation}
where
\begin{equation}
N_{G_3}=\frac{uvw}{(v^2v^2+u^2w^2+v^2w^2)^{1/2}}
\end{equation}
is a normalization coefficient.\\

 A plane can be parametrized by two parameters, say, $p$ and $q$.
Putting (without losing generality)
$$Y_{G_1}=p ~, ~~
%%\normalsize{and} ~~
Z_{G_1}= q $$
we get from (\ref{B.21})
\begin{equation}
G_1=
\left(\begin{array}{c}
X_{G_1}\\
Y_{G_1}\\
Z_{G_1}
\end{array}\right) =
N_{G_1}\left(\begin{array}{c}
-v[p/u + q/w]\\
p\\
q
\end{array}\right)
\label{B.25}
\end{equation}
where
\begin{equation}
N_{G_1}=\frac{1}{[v^2(p/u+q/w)^2 + p^2 + q^2]^{~1/2}}
\end{equation}

  The state $G_2$ orthogonal to $G_1$ and $G_3$ can be determined
(up to a sign) as the vector product
\begin{equation}
G_2= G_3 \times G_1
\end{equation}
written explicitly as
\begin{equation}
G_2=
\left(\begin{array}{c}
X_{G_2}\\
Y_{G_2}\\
Z_{G_2}
\end{array}\right) =
N_{G_2}\left(\begin{array}{c}
q/u -p/w\\
-v[p/u + q/w]/w - q/v\\
p/v + v[p/u + q/w]/u
\end{array}\right)
\label{B.28}
\end{equation}
where
\begin{equation}
N_{G_2}=N_{G_1}.N_{G_3}.
\end{equation}
Note that the role of $G_1$ and $G_2$ can be exchanged.\\

   Formulae (\ref{B.23}), (\ref{B.25}) and (\ref{B.28}) can be
combined in a unique one as follows
\begin{equation}
\left(\begin{array}{c}
G_1\\
G_2\\
G_3\\
\end{array}\right) =
\left( \begin{array}{ccc}
X_{G_1} & Y_{G_1} &Z_{G_1}\\
X_{G_2} & Y_{G_2} &Z_{G_2}\\
X_{G_3} & Y_{G_3} &Z_{G_3}
\end{array} \right)
\left( \begin{array}{c}
\zeta_\eta \\
\zeta_\rho\\
\zeta_\chi
\end{array} \right)
\label{B.30}
\end{equation}
where the matrix
$$
A=\left( \begin{array}{ccc}
X_{G_1} & Y_{G_1} &Z_{G_1}\\
X_{G_2} & Y_{G_2} &Z_{G_2}\\
X_{G_3} & Y_{G_3} &Z_{G_3}
\end{array} \right)\equiv (A^{-1})^T, ~~ \mbox{det}A=1,
$$
is an orthogonal matrix $SO(3)$ and has the explicit form

\begin{equation}
A = \left( \begin{array}{ccc}
-v[p/u+q/w] N_{G_1} & p N_{G_1} & q N_{G_1}\\[2mm]
[q/u-p/w] N_{G_2} & [-v(p/u+q/w)/w -q/v]N_{G_2}
& [v(p/u+q/w)/u +p/v]N_{G_2}\\[2mm]
N_{G_3}/v & N_{G_3}/u & N_{G_3}/w  \end{array} \right).
\label{B.31}
\end{equation}
Then, the relation
\begin{equation}
\left( \begin{array}{c}
\zeta_\eta \\
\zeta_\rho\\
\zeta_\chi
\end{array} \right)
=
\left( \begin{array}{ccc}
X_{G_1} &
X_{G_2} & X_{G_3}\\
Y_{G_1} &
Y_{G_2} & Y_{G_3}\\
Z_{G_1} &
Z_{G_2} & Z_{G_3}
\end{array} \right)
\left(\begin{array}{c}
G_1\\
G_2\\
G_3
\end{array}\right)
\end{equation}
is namely that inverse to (\ref{B.30}).\\

   As $w\gg u,v$, it is clear from Eq. (\ref{B.23}) that there is no way
to make the massive state $G_3$ parallel to the $\zeta_{\chi}$-- direction
(i.e., to impose the condition $X_{G_3}=Y_{G_3}=0 \neq Z_{G_3}$ is
impossible). It means that we always have $G_3$ different from
$\zeta_{\chi}$ which could be singled out from $G_1$ and $G_2$ only.
For example, in the limit $p\rightarrow 0$ and $w\rightarrow \infty ~~
(w\gg u, v)$ we should have
\begin{equation}
G_1=\left(\begin{array}{c}
0\\
0\\
1
\end{array}\right)\equiv \zeta_{\chi}~, ~~~~
G_2= \frac{1}{(u^2 + v^2)^{1/2}}\left(\begin{array}{c}
v\\
-u\\
0
\end{array}\right), ~~~~
G_3=\frac{1}{(u^2 + v^2)^{1/2}} \left(\begin{array}{c}
u\\
v\\
0
\end{array}\right).
\end{equation}
  In contradiction to Tonasse's results, $\zeta_{\chi}$ here is
not massive (as stated in \cite{ton}) but approximately massless,
while $\zeta_{\eta}$ and $\zeta_{\rho}$ are mixings between the
massless $G_2$ and the massive $G_3$:

\begin{equation}
\left(\begin{array}{c}
G_2\\
G_3\\
\end{array}\right) = \frac{1}{(v^2 + u^2)^{1/2}}
\left( \begin{array}{cc}
v & -u\\
u & v\end{array} \right)
\left(\begin{array}{c}
\zeta_{\eta} \\
\zeta_{\rho}\end{array} \right), ~~~~ G_1 = \zeta_{\chi}.
%\eqnum{B.21}
\end{equation}
The inverse relation is

\begin{equation}
\left(\begin{array}{c}
\zeta_{\eta}\\
\zeta_{\rho}\\
\end{array}\right) = \frac{1}{(v^2 + u^2)^{1/2}}
\left( \begin{array}{cc}
v & u\\
-u & v\end{array} \right)
\left(\begin{array}{c}
G_2\\
G_3 \end{array} \right), ~~~~ \zeta_{\chi} = G_1 .
%\eqnum{B.21}
\end{equation}

  The conclusion that $\zeta_{\chi}$ should be massless can be
intuitively seen from the following observation:
the mass terms associated with $\zeta_\eta,\  \zeta_\rho,\  \zeta_\chi$
fields in Higgs potential $V_T$, after imposing the constraints,
are
\begin{equation}
-\frac{f_1 u w}{2 v}, \ -\frac{f_1 v w}{ 2 u},\
-\frac{f_1 v u }{2 w},
%\eqnum{B.20}
\label{B.36}
\end{equation}
respectively. Therefore, in the limit  $w \gg v, u$ the last term  in
(\ref{B.36}) is smallest (massless).\par

  The reason why our results differ from those of Tonasse is that
Eqs. (10a) and (10b) in \cite{ton} do not represent relations between
orthogonal bases as they should have to be. We would say the same
for Eqs. (23a), (23b), (25a), (25b), (28a) and (28b) in that paper
\cite{ton} when the sextet is included.\par

  In the singly charged sector we have two Goldstone bosons and two
physical massive fields with mixings
\begin{equation}
\left(\begin{array}{c}
\eta^+_1\\
\rho^+\\
\end{array}\right) = \frac{1}{(v^2 + u^2)^{1/2}}
\left( \begin{array}{cc}
- v & u\\
u & v\end{array} \right)
\left(\begin{array}{c}
G^+_4 \\
H^+_5 \end{array} \right),
\label{B.37}
\end{equation}
\begin{equation}
\left(\begin{array}{c}
\eta_2^+\\
\chi^+\\
\end{array}\right) = \frac{1}{(v^2 + w^2)^{1/2}}
\left( \begin{array}{cc}
- v & w\\
w & v\end{array} \right)
\left(\begin{array}{c}
G^+_5 \\
H^+_6 \end{array} \right).
\label{B.38}
\end{equation}
The masses of $H^+_5$ and $H^+_6$ are given, respectively
\begin{equation}
m^2_{H^+_5} = \frac{v^2 + u^2}{2 v u}( f_1 w - 2 \lambda_7 v u), ~~
m^2_{H^+_6} = \frac{u^2 + w^2}{2 u w}( f_1 u - 2 \lambda_8 u w).
\label{B.39}
\end{equation}
Note, that at the considered (tree) level the mass spectrum and eigenstates 
in this sector are exact.\par

 The doubly charged sector contains one Goldstone boson $G^{++}$ and
one physical massive scalar $H^{++}$
\begin{equation}
\left(\begin{array}{c}
G^{++}\\
H^{++}\\
\end{array}\right) = \frac{1}{(u^2 + w^2)^{1/2}}
\left( \begin{array}{cc}
u & -w\\
w & u\end{array} \right)
\left(\begin{array}{c}
\rho^{++} \\
\chi^{++} \end{array} \right),
\label{B.40}
\end{equation}

\begin{equation}
\left(\begin{array}{c}
\rho^{++}\\
\chi^{++}\\
\end{array}\right) = \frac{1}{(u^2 + w^2)^{1/2}}
\left( \begin{array}{cc}
u & w\\
-w & u\end{array} \right)
\left(\begin{array}{c}
G^{++} \\
H^{++} \end{array} \right)
\label{B.41}
\end{equation}
with masses
\begin{equation}
m_{G^{++}} = 0, ~~
m^2_{H^{++}} = \frac{u^2 + w^2}{2 u w}( f_1 v - 2 \lambda_9 u w),
\label{B.42}
\end{equation}
respectively.\par

  Requiring that square masses of the physical fields are
positive (otherwise, they are Goldstone ones) and combining
Eqs.(\ref{B.4}), (\ref{B.6}), (\ref{B.8}), (\ref{B.13}), (\ref{B.39})
and (\ref{B.42}) we get the following relations between the parameters
of the potential
\begin{equation}
\frac{\lambda_1}{\lambda_2} \hspace*{0.2cm}  \left\{\begin{array}{c}
\hspace*{0.2cm}\stackrel{<}{\sim} u^4/ v^4,\  \mbox{if} \  v > u,\\
\approx 1 ,\ \hspace*{0.4cm} \mbox{if}\hspace*{0.2cm} \  v =  u,\\
\hspace*{0.2cm} \stackrel{>}{\sim} u^4 / v^4, \ \mbox{if} \  v < u,\\
\end{array}\right.
%\eqnum{B.24}
\label{B.43}
\end{equation}
\begin{equation}
\lambda_3  \stackrel{<}{\sim}  0,\hspace*{0.2cm} f_1\  < \
0;\hspace*{0.2cm}
\frac{f_1}{\lambda_7}\  <\  2
\frac{v u}{w} ,\hspace*{0.2cm}   \frac{f_1}{\lambda_8} < 2
\frac{v w}{u},\hspace*{0.2cm}  \frac{f_1}{\lambda_9} < 2
\frac{u w}{v}.
\label{B.44}
\end{equation}
Here the unnatural condition $v \neq u$ in~\cite{ton} is removed.\par


\begin{thebibliography}{99}
\bibitem{ppf} F. Pisano and V. Pleitez, Phys. Rev. {\bf D46},
410 (1992);\\
P. H. Frampton, Phys. Rev. Lett. {\bf 69}, 2889 (1992).
\bibitem{rf} R. Foot,
O. F. Hernandez, F. Pisano, and V. Pleitez, Phys. Rev. {\bf D47},
4158 (1993).
\bibitem{svs}M. Singer, J. W. F. Valle, and J. Schechter, Phys. Rev.
{\bf D22}, 738 (1980).
\bibitem{flt} R. Foot, H. N. Long, and Tuan A. Tran,
 Phys. Rev. {\bf D50}, R34 (1994).
\bibitem{hnl}H. N. Long, Phys. Rev. {\bf D54}, 4691 (1996).
\bibitem{mpp}J. C. Montero, F. Pisano, and V. Pleitez,
Phys. Rev. {\bf D47}, 2918 (1993).
\bibitem{pq}R. D. Peccei and H. R.  Quinn, Phys. Rev. Lett. {\bf 38},  1440
(1977);
Phys. Rev. {\bf D16}, 1791  (1977).
\bibitem{pal}P. B. Pal, Phys. Rev. {\bf D52}, 1659 (1995).
\bibitem{dng}D. Ng, Phys. Rev.  {\bf D49},  4805 (1994).
\bibitem{ton}M. D. Tonasse, Phys. Lett. {\bf B381}, 191 (1996).
\bibitem{gum}D. G. Dumm, Int. J. Mod. Phys. {\bf A11}, 887 (1996).
\bibitem{ic97}H. N. Long, Mod. Phys. Lett. 
{\bf A13}, 1865 (1998).
\bibitem{smir}V. I. Smirnov, {\it A course of higher mathematics},
vol. III, part one (Pergamon press, Oxford, 1964), pp 118 -- 125.
\bibitem{kor}G. A. Korn and T. M. Korn, {\it Mathematical
handbook for scientists and engineers}, (McGraw-Hill
Publishing Company, New York, 1968), pp 12 \& 407.
\end{thebibliography}
\end{document}